\documentstyle[12pt,twoside,epsfig,fleqn,espcrc1]{article}

\newcommand{\bra}{\langle}
\newcommand{\ket}{\rangle}

\title{The $\Xi^- +d \rightarrow n+\Lambda+\Lambda$ reaction 
        as a probe of the $\Lambda\Lambda$ interaction
        \thanks{The author would like to dedicate the present 
                contribution to the memory of Carl B. Dover.}}

\author{I. R. Afnan\address{Flinders University, South Australia}}

\begin{document}

\maketitle

\begin{abstract}
Within the framework of the Faddeev equations we demonstrate that
a $\Lambda\Lambda-\Xi N$ interaction that gives a $\Lambda\Lambda$ scattering
length comparable to the $nn$ scattering length, and the binding energy of
$^{\ \ 6}_{\Lambda\Lambda}$He as an $\alpha\Lambda\Lambda-\alpha\Xi N$ 
system, produces a final state interaction peak in the neutron spectrum
for the reaction $\Xi^- d \rightarrow n\Lambda\Lambda$. This suggests
that this reaction could be used to constrain the $\Lambda\Lambda$ 
scattering length.
\end{abstract}

\section{INTRODUCTION}

The main interest in the reaction $\Xi^- d \rightarrow n\Lambda\Lambda$
has been as a tool in the search for the $H$ dibaryon~\cite{B82,AD84},
(the strangeness $S=-2$ six-quark $SU(3)$ flavor singlet first suggested by 
Jaffe~\cite{J77}). On the other hand, the analogous reaction in
the $S=0$ channel (i.e. $n+d\rightarrow p+n+n$) has been used extensively to 
examine the final state interaction (FSI) between the two neutrons and the 
extraction of the $nn$ scattering length. This suggests that one could 
consider the reaction $\Xi^- d \rightarrow n\Lambda\Lambda$ as a means 
to study the $\Lambda\Lambda$ interaction. 

The success of the $n-d$ breakup reaction as a tool to extract the $nn$ 
scattering length is based on the fact that the $nn$ amplitude near zero 
energy is dominated by the $^1$S$_0$ anti-bound state pole. This pole in
the $nn$ amplitude generates a FSI peak in $n-d$ breakup that is sensitive 
to the $nn$ scattering length. To carry the same analysis for the 
$\Xi^- d \rightarrow n\Lambda\Lambda$, the $\Lambda\Lambda$ interaction 
in the $^1$S$_0$ should also be dominated by an anti-bound state. 

The only experimental data on the $\Lambda\Lambda$ interaction are the 
three observed $\Lambda\Lambda$ hypernuclei 
$^{\ \ 6}_{\Lambda\Lambda}$He~\cite{Pr66}, 
$^{\ 10}_{\Lambda\Lambda}$Be~\cite{Dan63,Dal89}, 
and $^{\ 13}_{\Lambda\Lambda}$B~\cite{Aok90,DMG91}, which invariably 
give an effective $S$-wave matrix element of
\begin{equation}
- \bra\,V_{\Lambda\Lambda}\,\ket \approx 4 - 5 \ \mbox{MeV}\ . \label{eq:1}
\end{equation}
This is smaller than the effective $nn$ matrix element $-\bra\,V_{nn}\,\ket
\approx 6-7$~MeV, but larger than the $\Lambda N$ $^1$S$_0$ matrix element
of $-\bra\,V_{\Lambda N}\,\ket\approx 2-3$~MeV. This has led to the 
suggestion by Dover~\cite{Do94} that the $\Lambda\Lambda$ system might 
support an anti-bound or weakly bound state considering the fact that the
$\Lambda$ mass is larger than the nucleon mass resulting in less kinetic 
energy in the $\Lambda\Lambda$ system, and the observation that the short range 
repulsion in the $nn$ interaction might not carry through to the $\Lambda
\Lambda$ system. There is also the additional attraction resulting from the
coupling between the $\Lambda\Lambda$ and $\Xi N$ channels which is
suppressed in $\Lambda\Lambda$ hypernuclei as a result of Pauli 
blocking~\cite{G94}.

With the absence of any $S=-2$ two-body data, the simplest procedure to
construct a baryon-baryon interaction is to resort to a flavor $SU(3)$ 
rotation of the one-boson-exchange (OBE) potential from the $S=0, -1$ to the
$S=-2$ channel. This allows us to determine all the meson-baryon 
coupling constants in the $S=-2$ channel. The $SU(3)$ breaking is  
then partly due to the fact that the masses of the baryon and meson are 
taken from experiment, and partly as a result of modifying the short range 
interaction in the different $S$ channels.

To qualitatively see how the effective $\Lambda\Lambda$ interaction 
could possibly be comparable in strength to the $nn$ potential, we first 
write the diagonal elements of the potential in the particle basis in 
terms of the flavor symmetric $\{8\otimes 8\}$ irreducible representation 
of $SU(3)$~\cite{deS63}, i.e.,
\begin{eqnarray}
\bra nn|V|nn\ket &\equiv&\bra\,V\,\ket_{nn} = V_{27} \nonumber \\
\bra \Lambda N|V|\Lambda N\ket &\equiv& \bra\,V\,\ket_{\Lambda N} =
         \frac{36}{40}\,V_{27} + \frac{4}{40}\,V_{8_s} \nonumber \\
\bra\Lambda\Lambda|V|\Lambda\Lambda\ket &\equiv&\bra\,V\,\ket_{\Lambda\Lambda}
        =  \frac{27}{40}\,V_{27} + \frac{8}{40}\,V_{8_s} 
        + \frac{5}{40}\,V_{1}\ .              \label{eq:2}
\end{eqnarray}
Since the $nn$ interaction in the $^1$S$_0$ is pure $\{27\}$ representation
and is strong, we would expect the $V_{27}$ to give the dominant contribution
for all three interactions listed above. In fact, for the Nijmegen soft core
potential, the $V_{27}$ representation dominates the medium to long range 
part of the interaction, while $V_{1}$ has almost zero contribution for 
$r>1$~fm~\cite{RM92}. We therefore have what is expected~\cite{Do94}, i.e.,
\begin{equation}
\bra\ V_{nn}\ \ket > \bra\ V_{\Lambda N}\ \ket 
> \bra\ V_{\Lambda\Lambda}\ \ket\ .          \label{eq:3}
\end{equation}
The results in Eq.~(\ref{eq:2}) are the lowest order contribution to the
diagonal amplitudes in the $S=0,-1,-2$ channels. The fact that in the 
$^1$S$_0$ both the $\Lambda N$ and the $\Lambda\Lambda$ interaction are 
part of a coupled channel problem, suggests that the coupling to lowest 
order gives further attraction to the amplitude, and this attraction  
is of the form
\begin{eqnarray}
V_{\Lambda N}^{\rm eff} &\approx& \bra V\ket_{\Lambda N} - 
                   \frac{|\bra \Lambda N|V|\Sigma N\ket|^2}{\Delta E_{YN}}
                   \quad\mbox{where}\quad \ 
                   \Delta E_{YN}\approx 80\,\mbox{MeV}\nonumber\\
V_{\Lambda\Lambda}^{\rm eff} &\approx& \bra V\ket_{\Lambda\Lambda} - 
                   \frac{|\bra\Lambda\Lambda |V|\Xi N\ket|^2}
                        {\Delta E_{\Lambda\Lambda}}\quad\mbox{where} 
             \quad\Delta E_{\Lambda\Lambda}\approx 25\,\mbox{MeV}\label{eq:4}
\end{eqnarray}
where~\cite{deS63}
\begin{eqnarray}
\bra\Lambda N|V|\Sigma N\ket &=& -\frac{12}{40}\,V_{27} 
                                 + \frac{12}{40}\,V_{8_s} \nonumber \\
\bra\Lambda\Lambda|V|\Xi N\ket &=& - \frac{18}{40}\,V_{27} 
                                 + \frac{8}{40}\,V_{8_s}
                                 + \frac{10}{40}\,V_1\ .  \label{eq:5}
\end{eqnarray}
From Eqs.(\ref{eq:4}) and (\ref{eq:5}), it is clear that if the $\{27\}$
is the dominant contribution to the flavor symmetric $SU(3)$ representation
of $\{8\otimes 8\}$, then the coupling in the $S=-2$ channel is more 
important that that in the
$S=-1$ channel, and it is possible that the $\Lambda\Lambda$ amplitude could 
give a scattering length that is comparable to the $nn$ scattering length, 
and larger in magnitude than the scattering length in the $\Lambda N$ 
channel.

In Sec.~2 we briefly describe the construction of the $S=-2$
potential~\cite{C96,CAG97} corresponding to the $SU(3)$ rotation of the 
Nijmegen model $D$ potential~\cite{NRS77}. Here the short range part of the 
interaction is chosen such that the $^1$S$_0$ $I=0$ $S=-2$ system has no bound 
state, an anti-bound state or a bound state. We then proceed 
in Sec.~3 to present the binding energy of $^{\ \ 6}_{\Lambda\Lambda}$He 
as a $\alpha\Lambda\Lambda-\alpha\Xi N$ three-body model for the different 
$S=-2$ potentials~\cite{C96,CAG97}. Here we find that the potential with a
$\Lambda\Lambda$ anti-bound state gives a binding energy for
$^{\ \ 6}_{\Lambda\Lambda}$He closest to the experimental 
result~\cite{Pr66}. In Sec.~4 we turn to the reaction 
$\Xi d\rightarrow n\Lambda\Lambda$ and examine the sensitivity of 
the final state interaction peak to the $\Lambda\Lambda$ scattering 
length~\cite{C96,CAG98}. Finally, in Sec.~5 we present some concluding remarks.

\section{THE $S=-2$ $B-B$ POTENTIALS}

In the absence of any data in the $S=-2$ channel, we consider the
meson exchange part of the Nijmegen model $D$ potential for the $NN$ and  
$YN$ systems~\cite{NRS77}, and perform an $SU(3)$ rotation to determine the 
coupling constants of the mesons to the baryons. For a purely $S$-wave 
interaction, the resultant OBE potential for the exchange of the $i^{\rm th}$ 
meson is given by
\begin{equation}
V_i(r) = V^{(i)}_c(r) 
   + \vec{\sigma}_1\cdot\vec{\sigma}_2\ V^{(i)}_\sigma(r)\ .\label{eq:6}
\end{equation}
\begin{figure}[h]
\begin{minipage}[h]{.49\linewidth}
  \centering\epsfig{figure=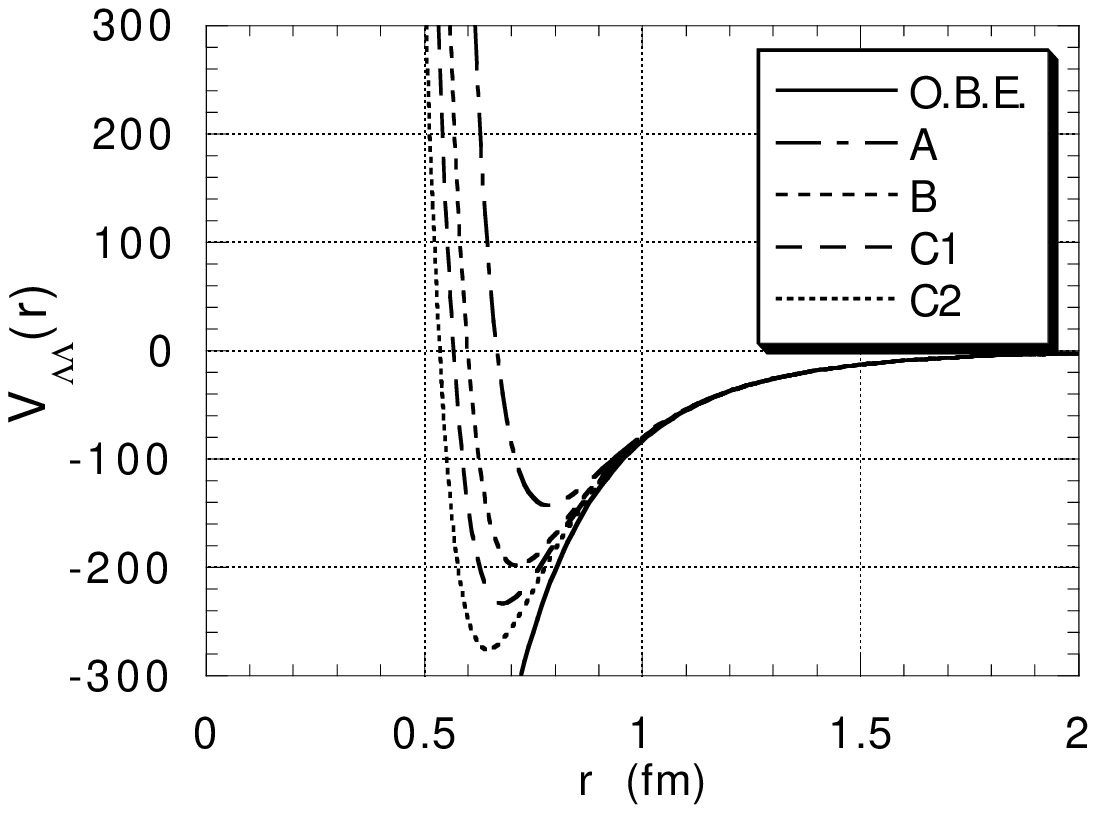,width=\linewidth}
\end{minipage}\hfill
\begin{minipage}[h]{.49\linewidth}
  \centering\epsfig{figure=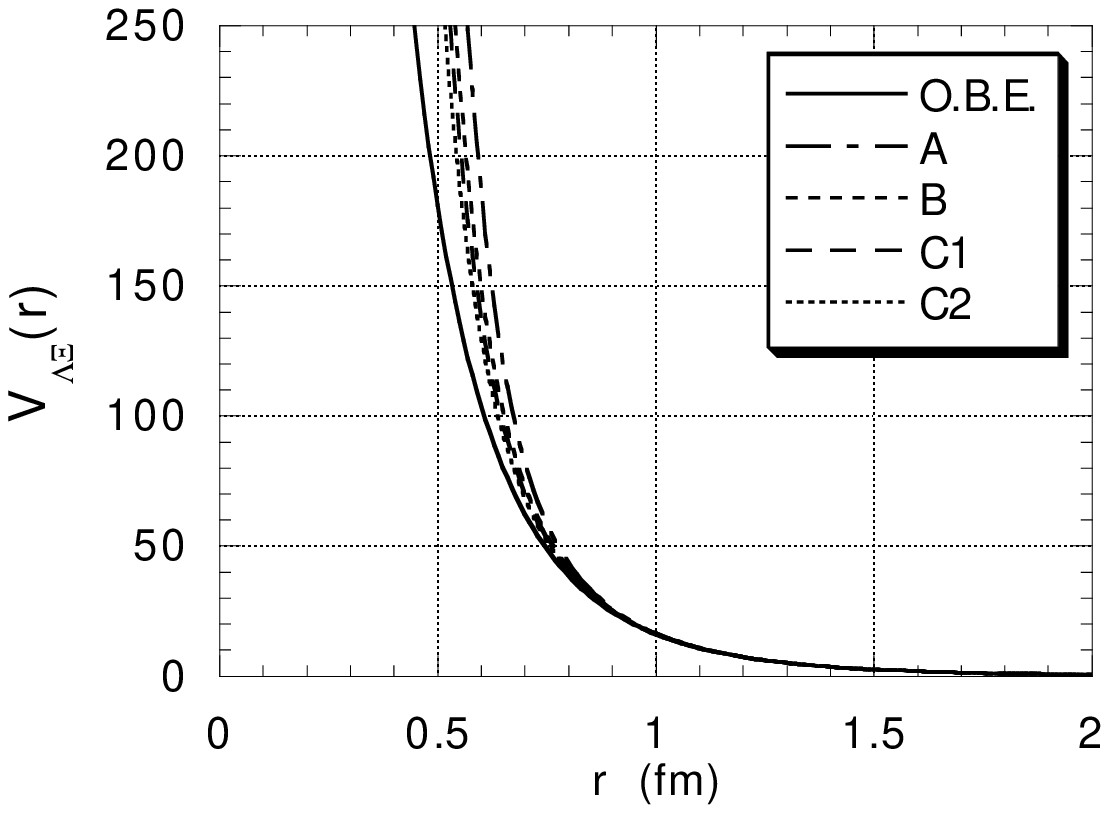,width=\linewidth}
\end{minipage}
\caption{The strangeness $-2$, $^1$S$_0$, one-boson-exchange potentials 
$V_{\Lambda\Lambda}$ and $V_{\Lambda\Xi}$.}\label{fig.1}
\end{figure}

Since the resulting OBE potential is singular at the origin, we 
introduce a repulsive soft core with a cut-off mass 
$M\approx 2.5$~GeV. As a result, the radial potential for the 
exchange of the $i^{\rm th}$ meson is:
\begin{equation}
V_{\alpha}^{(i)}(r) = V_0^{(i)} \left[\frac{e^{-m_ir}}{m_ir} 
          - C\,\left(\frac{M}{m_i}\right)\, 
                     \frac{e^{-M r}}{M r}\right]
            \qquad \alpha = c,\sigma             \ ,\label{eq:7}
\end{equation}
where $m_{i}$ is the mass of the exchanged meson, and $V_{0}^{(i)}$ 
is given in terms of the masses and the coupling constants as determined 
by the $NN$ and $YN$ data \cite{C96,CAG97}. The cut-off parameters $M$ 
and $C$ are now adjusted to ensure that the long range part 
$(r>0.8\ {\rm fm})$ of the meson exchange potential is not modified 
(see Fig.~\ref{fig.1}). The final parameters $M$ and $C$ in the
$\Lambda\Lambda$--$\Xi N$ interaction have been chosen to either support a 
bound state (C), generate an anti-bound state (B), or have no bound states 
at all (A). This allows us to test the hypothesis that the $\Lambda\Lambda$
amplitude in the $^1$S$_0$ interaction is comparable in 
strength to the $^1$S$_0$ $nn$ amplitude \cite{Do94,Dover}.

In Fig.~\ref{fig.1} we present the $\Lambda\Lambda$ potential 
$V_{\Lambda\Lambda}$ and and the potential for the coupling 
$\Lambda\Lambda-\Xi N$ $V_{\Lambda\Xi}$ in the $^1$S$_0$ isospin zero  
channel. Also included is the OBE potential with no cut-off.  Here we note 
that although the coupling potential $V_{\Lambda\Xi}$ is smaller in 
magnitude than $V_{\Lambda\Lambda}$, it is large enough to support the 
argument stated in the Introduction on the importance of the
coupling when considering the analysis of the data from $\Lambda\Lambda$
hypernuclei, and the extraction of effective $\Lambda\Lambda$ $S$-wave 
matrix elements. 

The above $SU(3)$ rotation and cut-off procedure gives a baryon-baryon 
potential in the $S=-2$ channel that is local in coordinate space. 
This potential is transformed into momentum space to solve the 
Lippmann-Schwinger equation. To reduce the Faddeev equations for
$^{\ \ 6}_{\Lambda\Lambda}$He and $\Xi d\rightarrow\Lambda\Lambda N$ systems 
from two-dimensional integral equations to a set of coupled one dimensional 
integral equations, separable potentials were constructed that give the 
same effective range parameters as the original OBE potentials in all 
$S$-waves~\cite{C96,CAG97}. In the introduction
of the cut-off and the construction of the separable potential, great care
has been taken to maintain the relative strength of the potential in the 
different spin-isospin channels, and in this way we retain the features of
the potentials resulting from the $SU(3)$ rotation of the original Nijmegen
model $D$ potential.

\section{BINDING ENERGY OF $_{\Lambda\Lambda}^{\ \; 6}$He}

To test the resulting potentials with the only experimental data on 
the $\Lambda\Lambda$ interaction (i.e. $\Lambda\Lambda$ hypernuclei), we have
chosen the lightest of the $S=-2$ hypernuclei for which we can construct
a reasonable three-body model, i.e., $_{\Lambda\Lambda}^{\ \; 6}$He. 
If we maintain the coupling between the $\Lambda\Lambda$ and $\Xi N$ 
channels, then $_{\Lambda\Lambda}^{\ \; 6}$He may be modeled as the two 
channel three-body system $\alpha\Lambda\Lambda-\alpha\Xi N$. The
three-body Alt Grassberger Sandhas (AGS) equations \cite{AGS67} can
now be solved exactly. The main sources of error in this model are:
(i) The need to model the Pauli blocking between the $N$ and the $\alpha$
in the $\alpha\Xi N$ channel. This is achieved by introducing a repulsive
$S$-wave interaction as has been implemented in $^6$Li as an $\alpha NN$ 
system~\cite{EA92}.
(ii) The $\alpha$ particle is taken to be elementary, i.e., we have not 
included the $\alpha^*\Lambda\Lambda$ channel, even though the energy of 
such a state is comparable to the energy of the $\alpha\Xi N$ state. 
This is partly a result of the fact that we do not have sufficient data
to fix the additional parameters introduced as a result of the coupling 
of the $\alpha N$ to the $\alpha^* N$ channel, and partly due to the 
observation that the coupling to the $\alpha\Xi N$ channel is suppressed 
due to Pauli blocking. 

\begin{table}[h]
\caption{The binding energy in MeV of $^{\ \; 6}_{\Lambda\Lambda}$He for the four
  potentials under consideration. Also included are the $\Lambda\Lambda$ 
scattering length $a_{\Lambda\Lambda}$ and binding energy.}\label{table.1}
\begin{tabular}{l|cccc|c} \hline
              & SA    & SB    & SC1    & SC2    & Exp.\cite{Pr66} \\ \hline
$a_{\Lambda\Lambda}$ in fm    & -1.90 & -21.0 & 7.84 & 3.36 \\ 
B.E.($\Lambda\Lambda$) in MeV & UB  & UB   & 0.71 & 4.74 \\ \hline
$\alpha\Lambda\Lambda$ -- $\alpha\Xi N$ & 
9.738 & 12.268 & 15.912 & 19.836 & 10.9$\pm$ 0.8 \\
$\alpha\Lambda\Lambda$ with no coupling to $\alpha\Xi N$ &
9.508 & 11.606 & 14.533 & 17.508 &  \\
$\alpha\Lambda\Lambda$ with effective $\Lambda\Lambda$ potential & 
10.007 & 14.134 & 17.842 & 23.750 & \\ \hline 
\end{tabular}
\end{table}

To examine the role of the coupling in the $\Lambda\Lambda$--$\Xi N$ 
channels, we have performed three distinct calculations by: 
(i)~Including the coupling between the two channels, and solving the 
equations for the $\alpha\Lambda\Lambda$--$\alpha\Xi N$ system. 
(ii)~Discarding the coupling between the channels at the amplitude 
level without any modification to the parameters of the potential.
This reduces the problem to the $\alpha\Lambda\Lambda$ three-body problem. 
(iii)~Excluding the coupling between the channels at the two-body level, 
but then adjusting the parameters of the potential to give the same 
$\Lambda\Lambda$ effective range parameters as the corresponding local 
OBE potential. The results for the different two-body interactions and 
different approximations are presented in Table~2 where we have also 
included the $\Lambda\Lambda$ scattering length $a_{\Lambda\Lambda}$ 
and the $\Lambda\Lambda$ binding energy for the potentials that support 
a bound state.

If we compare the binding energies of $_{\Lambda\Lambda}^{\ \; 6}$He 
for the different potentials with the one experimental measurement 
of $10.9\pm0.8$~MeV~\cite{Pr66} (see Table~\ref{table.1}), we observe that:
(i)~The potential SB, in the complete calculation (i.e. row three of
Table~\ref{table.1}), predicts the result closest to the experimental 
binding energy, and therefore is the best representation for the 
$\Lambda\Lambda$ interaction. This supports the suggestion that the 
$\Lambda\Lambda$ $^1$S$_0$ amplitude may in fact be comparable to that of the 
$nn$ $^1$S$_0$ amplitude, i.e., the $\Lambda\Lambda$ scattering length
is comparable to the $nn$ scattering length.
(ii)~A comparison of rows three and four of Table~1 demonstrates that 
the contribution of the coupling between the $\Lambda\Lambda$ and $\Xi N$ in 
$^{\ \; 6}_{\Lambda\Lambda}$He is small. This is due to the fact that 
the nucleon in the $\alpha\Xi N$ Hilbert space is Pauli blocked. 
(iii)~A comparison of rows three and five of Table~\ref{table.1}, suggests
that the inclusion of the coupling at the two-body level is essential if we
are to avoid over-binding the $\Lambda\Lambda$ hypernuclei nuclei.

\begin{figure}[h]
\begin{minipage}[b]{.49\linewidth}
  \centering\epsfig{figure=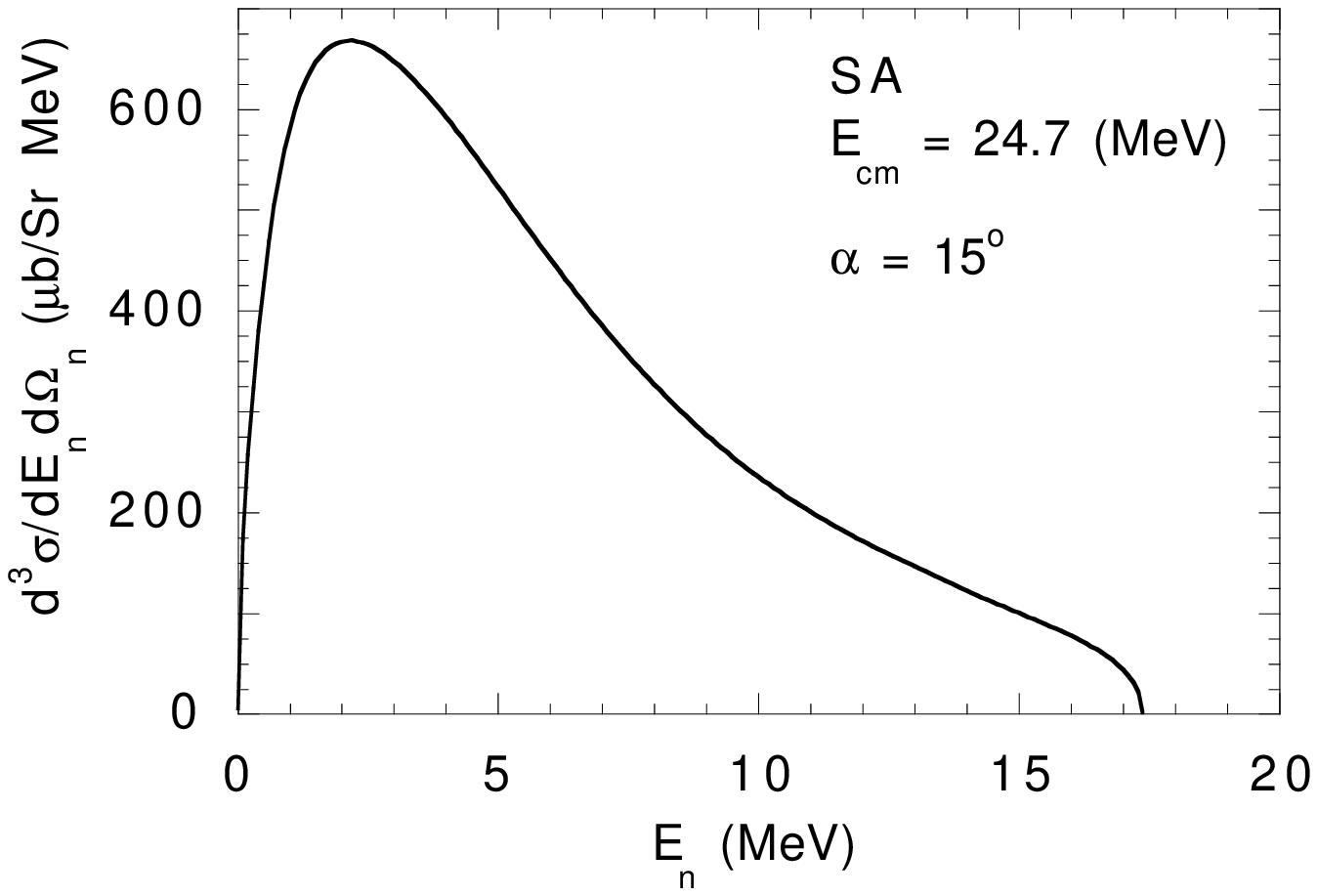,width=\linewidth}
\end{minipage}\hfill
\begin{minipage}[b]{.49\linewidth}
  \centering\epsfig{figure=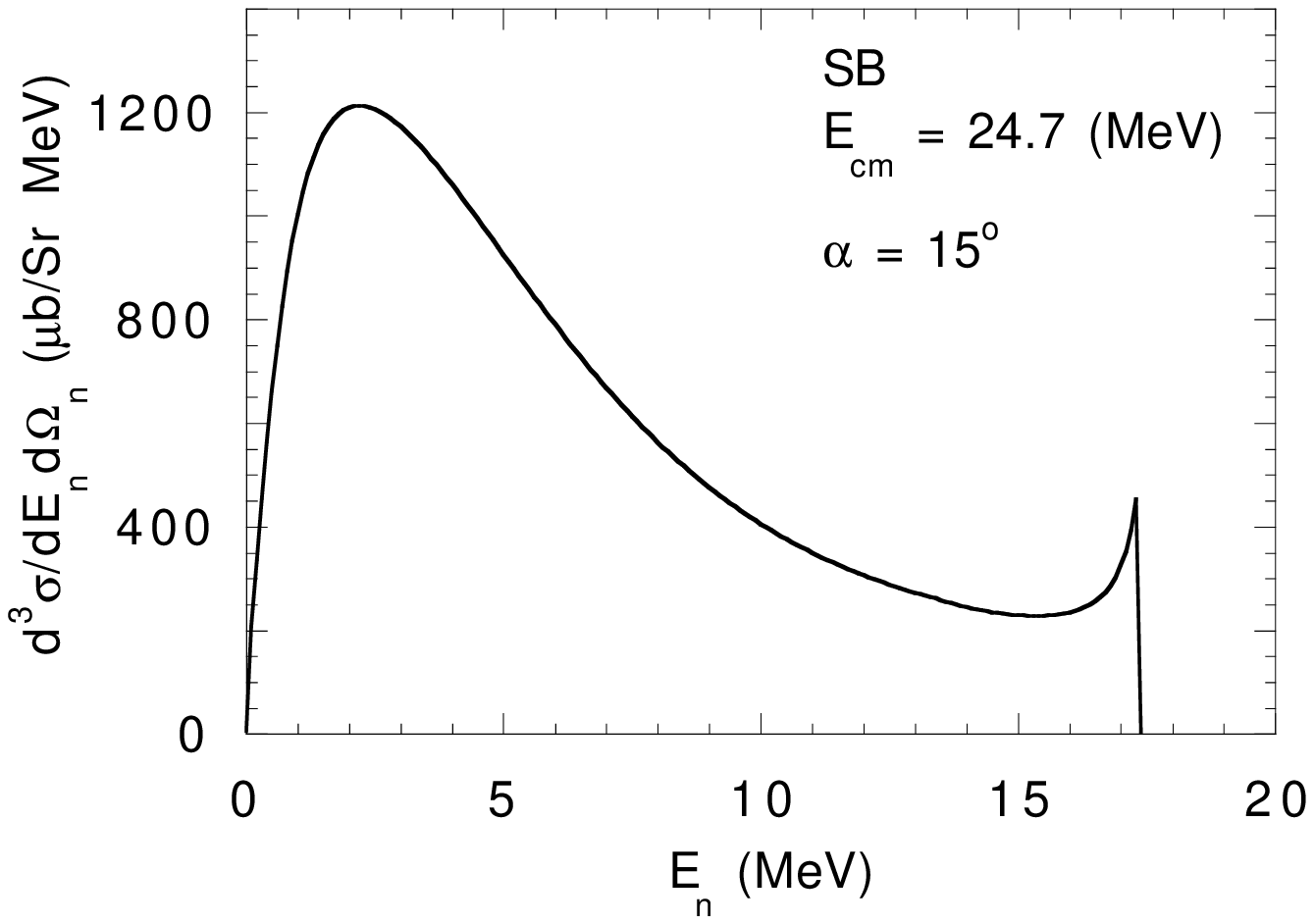,width=\linewidth}
\end{minipage}
\caption{The NDES for the potentials SA and SB.}\label{fig.2}
\end{figure}

\begin{figure}[h]
\begin{minipage}[b]{.49\linewidth}
  \centering\epsfig{figure=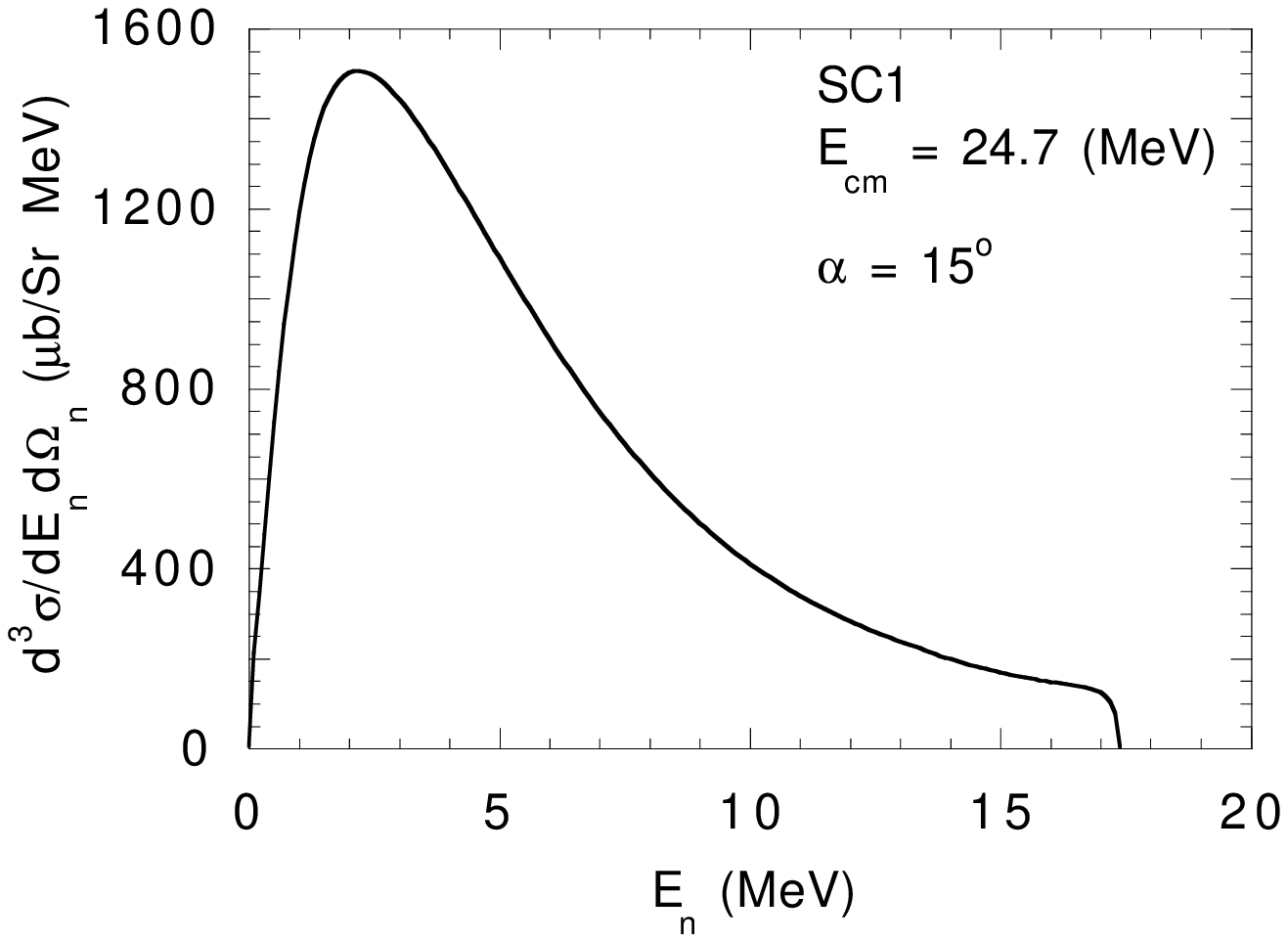,width=\linewidth}
\end{minipage}\hfill
\begin{minipage}[b]{.49\linewidth}
  \centering\epsfig{figure=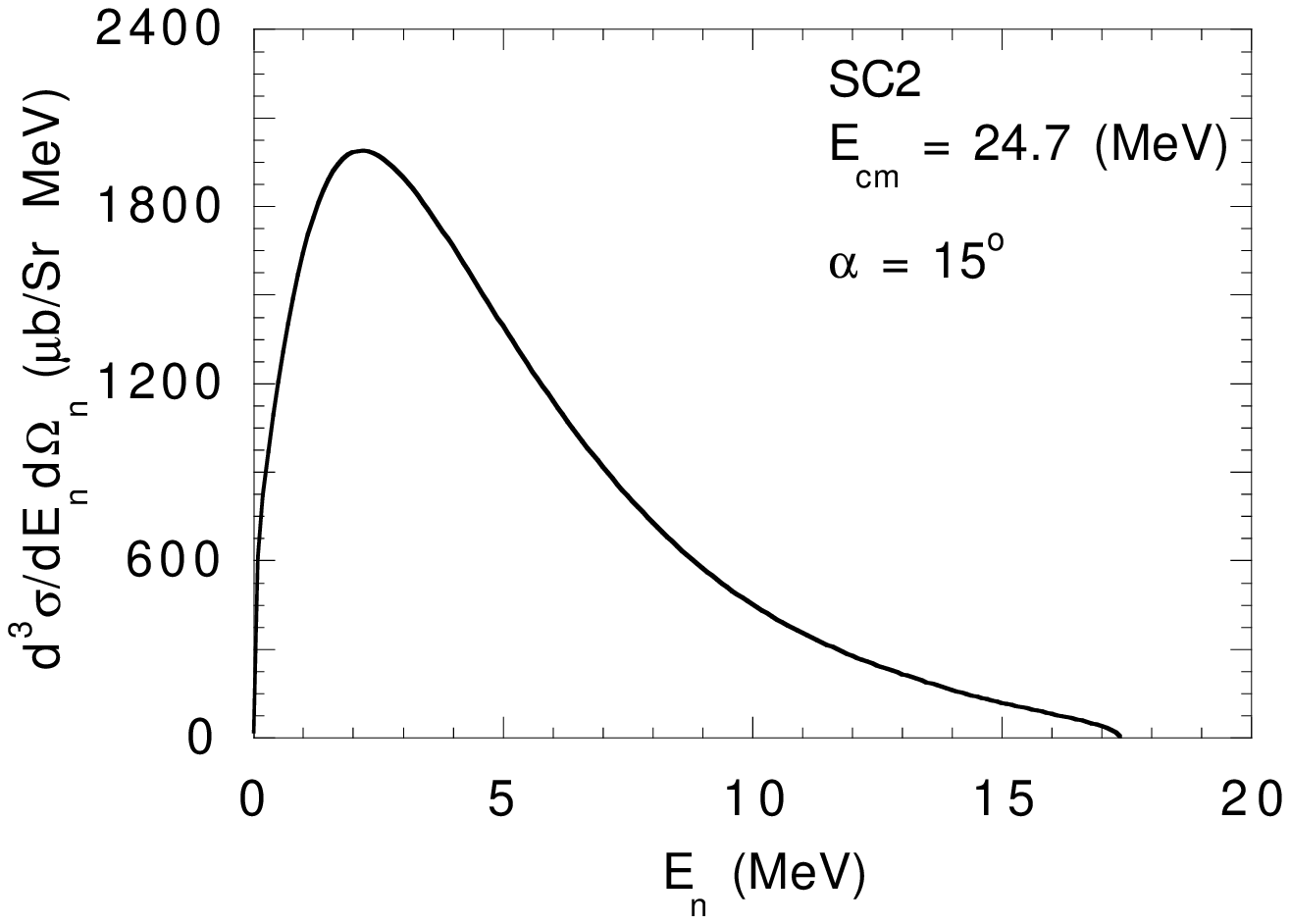,width=\linewidth}
\end{minipage}
\caption{The NDES for the potentials  SC1, and SC2.}\label{fig.3}
\end{figure}

\section{THE FINAL STATE INTERACTION IN $\Xi^-d\rightarrow n\Lambda\Lambda$}

We now turn to the reaction $\Xi^-d\rightarrow n\Lambda\Lambda$ for
which there is an experiment in progress at Brookhaven \cite{B92,M97}. In the 
Figs.~\ref{fig.2} and \ref{fig.3}  we show the neutron differential 
energy spectrum (NDES) for this reaction for the four 
$\Lambda\Lambda-\Xi N$ $^1$S$_0$ potentials under consideration. The energy 
at which the calculations have been performed corresponds to an incident $\Xi^-$ 
with an energy of 1~MeV (24.7~MeV relative to the $n\Lambda\Lambda$ threshold). 
This is an  approximation to the experimental setup in which the $\Xi^-$ 
is captured by the deuteron. In this way we avoid the complication of 
introducing an initial state Coulomb interaction into the three-body 
calculation. With the exception of the result for the potential 
SB, the neutron spectra do not exhibit the FSI peak expected. 
This suggests that we may use this reaction to determine the 
$\Lambda\Lambda$ scattering length. In all four cross sections the 
dominant feature is the large broad peak at the low-energy 
end of the neutron spectrum. 

\begin{figure}[h]
\centering\epsfig{figure=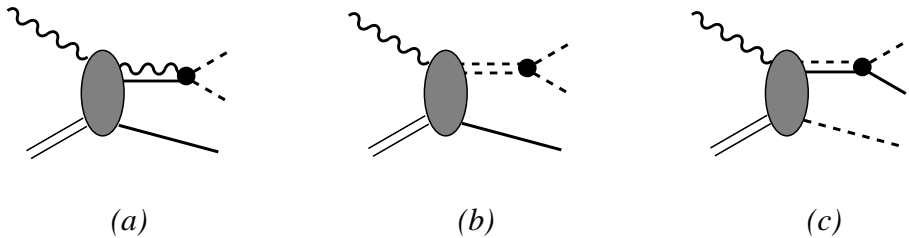,width=12cm}
\caption{The three amplitudes that contribute to the NDES for 
$\Xi^- d\rightarrow n\Lambda\Lambda$.}\label{fig.4}
\end{figure}

In Fig.~\ref{fig.4} we give a diagrammatic representation of the three
amplitudes that contribute to the cross section for 
$\Xi^- d\rightarrow n\Lambda\Lambda$. Diagrams (a) and (b)
are expected to contribute to the FSI peak, since the
final interaction is in the $\Lambda\Lambda-\Xi N$ coupled channels
which is dominated by the anti-bound state pole for the potential SB . 
On the other hand, diagram (c) is a background term that could
interfere constructively with either or both of the amplitudes corresponding 
to diagrams (a) and (b).

\begin{figure}[ht]
\begin{minipage}[b]{.49\linewidth}
  \centering\epsfig{figure=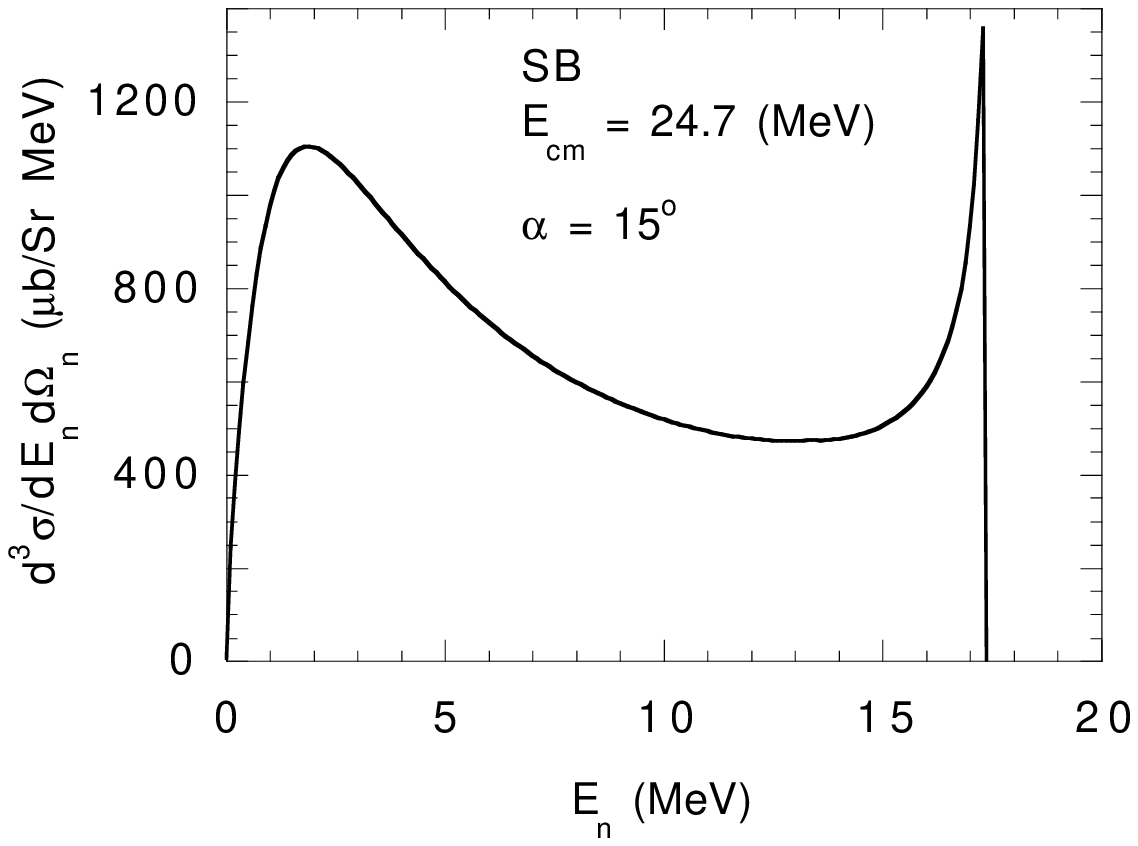,width=\linewidth}
  \caption{The NDES for diagram (a). }\label{fig.5}
\end{minipage}\hfill
\begin{minipage}[b]{.49\linewidth}
  \centering\epsfig{figure=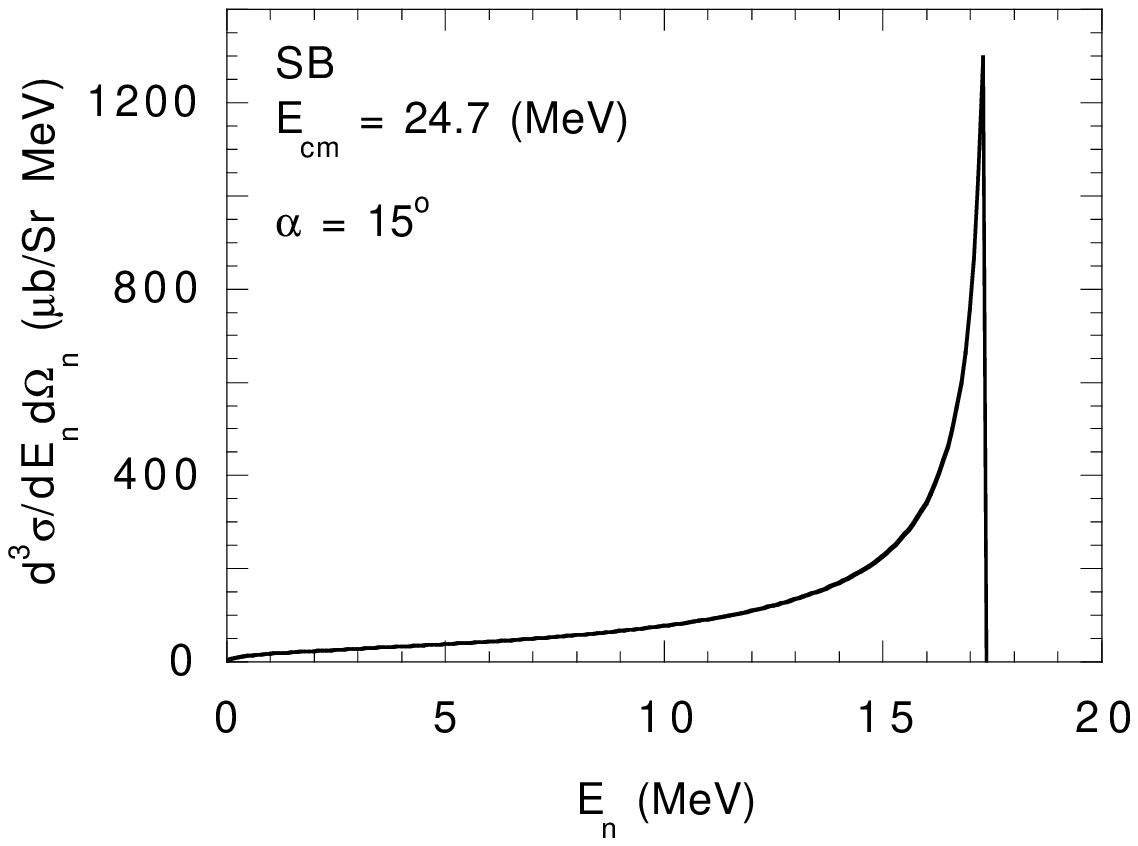,width=\linewidth}
  \caption{The NDES for diagram (b).}\label{fig.6}
\end{minipage}
\end{figure}

A detailed investigation of the different contributions to the NDES 
reveals that the suppression of the FSI is the result of a destructive 
interference between the amplitudes that contribute to the NDES. To
first establish that diagrams $(a)$ and $(b)$ have equal contribution
from the $\Lambda\Lambda$ anti-bound state, we present in Figs.~\ref{fig.5} 
and \ref{fig.6} the NDES resulting from diagrams $(a)$ and $(b)$ 
respectively. Here we observe that the magnitude of the FSI peak is almost 
identical for the two diagrams, which is expected considering the fact that 
the same anti-bound state pole dominates both amplitudes in the FSI region. 
Furthermore, the cross section in the FSI region is substantially larger 
than the cross section resulting from including all three diagrams in 
Fig.~\ref{fig.4}. 

The major difference between the two cross sections for diagrams $(a)$ and
$(b)$ in Fig.~\ref{fig.4} is at the low neutron energy end of the spectrum. 
Here the contribution from diagram $(a)$ is substantial, while the 
contribution of diagram (b) is small. To understand this difference, 
we observe that the lowest order approximation to diagram $(a)$ involves 
the $\Xi^-$ interacting with the proton in the deuteron to generate the 
two final $\Lambda$ hyprons. As a result, the neutron spectrum is directly 
related to the momentum distribution of the neutron in the deuteron. In 
other words, the peak in the neutron energy spectrum at low neutron energies 
in Fig.~\ref{fig.5} is a measure of the momentum distribution of the 
neutron in the deuteron, which in this case is taken to be that resulting 
from a rank one Yamaguchi separable potential. On the other hand, the 
lowest order contribution to diagram $(b)$ involves the $\Xi^-$ interacting 
with the proton in the deuteron converting the $\Xi^- p$ to two $\Lambda$ 
hyprons, but now one of the $\Lambda$ hyprons needs to rescatter off the 
neutron before we have a final state $\Lambda\Lambda$ interaction. In other 
words, the lowest order contribution to the amplitude representing diagram 
$(b)$ is third order in the multiple scattering series.  This has the effect 
of distorting the momentum distribution of the neutron in the deuteron. 
The fact that the multiple scattering series for 
$\Xi^- d\rightarrow (\Lambda\Lambda)n$ and $\Xi^- d\rightarrow (\Xi^- p)n$
do not converge, suggests that the neutron momentum distribution in the
deuteron in diagram $(b)$ gets completely smeared in the NDES in 
Fig.~\ref{fig.6}.

\begin{figure}[ht]
\begin{minipage}[b]{.49\linewidth}
  \centering\epsfig{figure=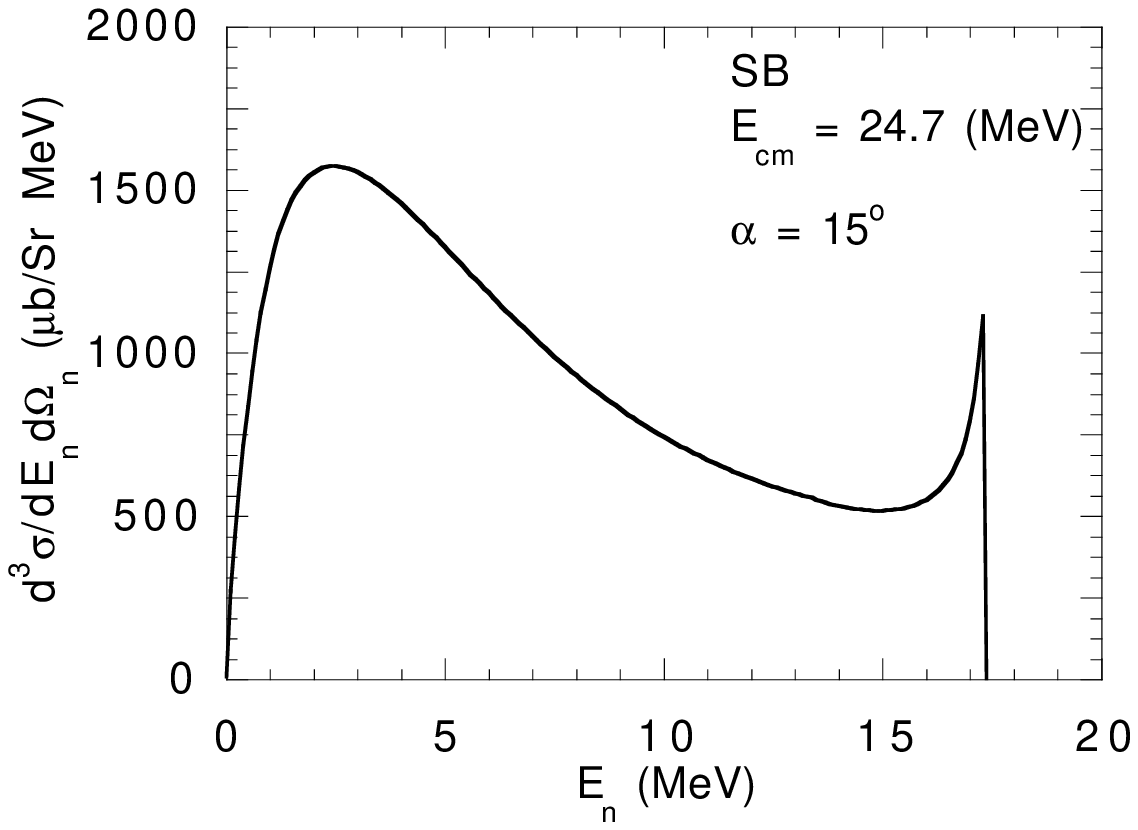,width=\linewidth}
  \caption{The NDES for (a) plus (c). }\label{fig.7}
\end{minipage}\hfill
\begin{minipage}[b]{.49\linewidth}
  \centering\epsfig{figure=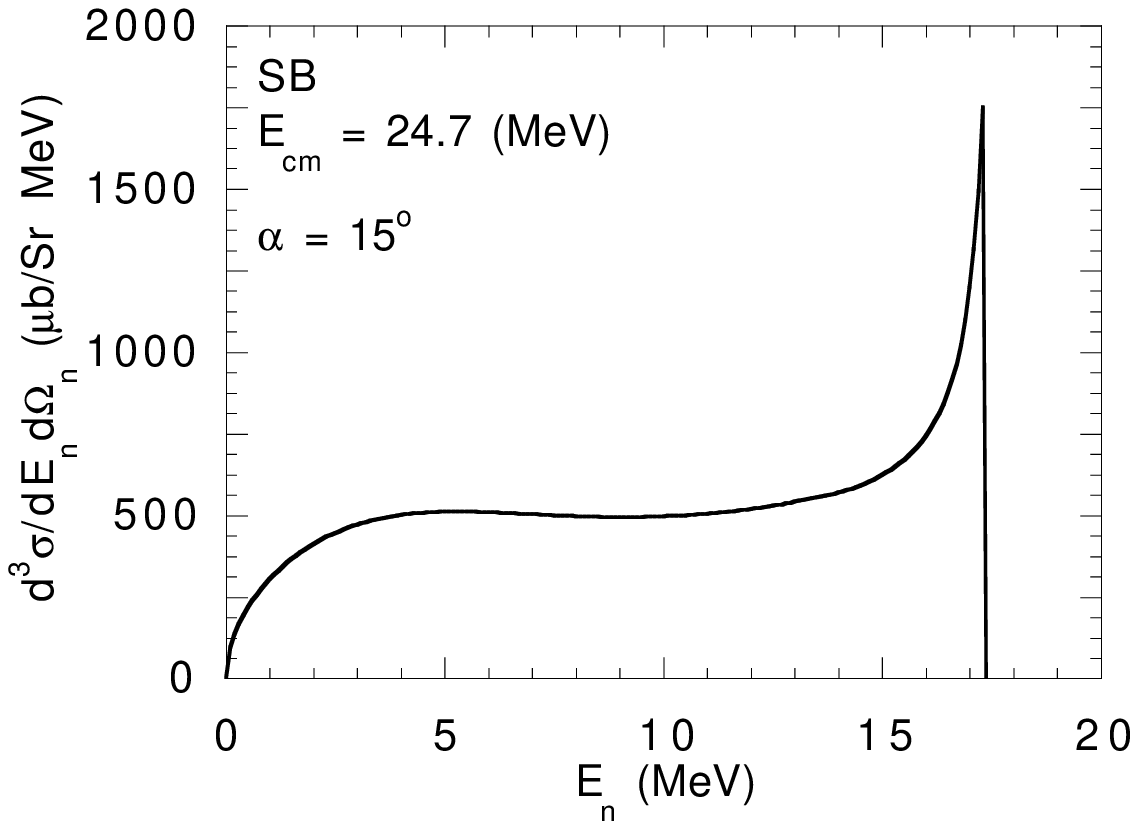,width=\linewidth}
  \caption{The NDES for (b) plus (c).}\label{fig.8}
\end{minipage}
\end{figure}

To determine the relative sign of the three amplitudes, we present in
Figs.~\ref{fig.7} and \ref{fig.8} the NDES for the diagrams (a) plus (c) 
and (b) plus (c) respectively. Here from the magnitude of the height 
of the FSI peak, we may conclude that diagrams (a) and (c) interfere 
destructively in the FSI region, while diagrams (b) and (c) give an 
enhancement to the FSI peak. This implies that diagrams (a) and (b) 
are out of phase. Since both of these diagrams are dominated by the 
anti-bound state in the FSI region, the fact that they are out of phase 
implies that in the cross section the FSI peak is suppressed. This is 
to be compared with $n-d$ breakup, where the final state interaction 
(i.e. $nn$) is not a coupled channel. In that case there is only one 
amplitude that is dominant in the FSI region, and the 
interference between this single dominant amplitude and the background 
amplitude narrows the peak in the proton spectrum. However, in 
$\Xi^- d\rightarrow n\Lambda\Lambda$ where the major interference is 
between two amplitudes dominated by the anti-bound state pole, the 
FSI peak should be more sensitive to the $\Lambda\Lambda$ scattering
length. This is achieved at a cost of a reduction in the magnitude of 
the FSI peak. This suggests that the reaction 
$\Xi^- d\rightarrow n\Lambda\Lambda$ could be
a means for determining the $\Lambda\Lambda$ scattering length, and in
this way directly establish that the $nn$ and $\Lambda\Lambda$ 
amplitudes at threshold are comparable in magnitude, as first suggested by 
Dover~\cite{Do94}.

\section{CONCLUSIONS}

The above analysis is based on a separable approximation to the $S$-wave OBE 
potentials that results from the $SU(3)$ rotation to the $S=-2$ channel 
of the Nijmegen model $D$ potential. Here we find that: 
(i)~A potential that gives a $\Lambda\Lambda$ scattering length that is 
comparable to the $nn$ scattering length in the $^1$S$_0$ partial wave, 
gives a binding energy of 12.27~MeV for $^{\ \ 6}_{\Lambda\Lambda}$He. This is 
close to the experimental binding energy of $10.9\pm0.8$~MeV~\cite{Pr66},
suggesting that the $\Lambda\Lambda$ and $nn$ amplitudes are
comparable at threshold.
(ii)~This same potential (SB) that gives a binding energy for 
$^{\ \ 6}_{\Lambda\Lambda}$He comparable to the experimental results, 
predicts a FSI peak for $\Xi d\rightarrow n\Lambda\Lambda$. 
This FSI peak is sensitive to the $\Lambda\Lambda$ scattering length as
a result of destructive interference between two amplitudes dominated by
the $\Lambda\Lambda$ anti-bound state pole. The fact that there are two
amplitudes that are dominant in the FSI region is a direct result of
the coupling between the $\Lambda\Lambda$ and $\Xi N$ channels. This
is to be compared with $n-d$ breakup in which only one amplitude is
dominated by the $nn$ anti-bound state pole. As a result, the FSI peak
in $\Xi^-d\rightarrow n\Lambda\Lambda$ could place a constraint on
the $\Lambda\Lambda$ scattering length if there is an anti-bound state
or very weakly bound state in the $\Lambda\Lambda$ system. 
(iii)~The low energy part of the neutron spectrum seems to be dominated
by the momentum distribution of the neutron in the deuteron, and a more
realistic deuteron wave function could enhance the magnitude of the 
cross section in the final state interaction region. 

From the above results we may deduce that a good measurement of the
NDES for $\Xi^- d\rightarrow n\Lambda\Lambda$ could be used to directly 
constrain the $\Lambda\Lambda$ scattering length, and in this way avoid 
the more complex procedure of extracting the $\Lambda\Lambda$ interaction 
from $\Lambda\Lambda$-hypernuclei.

\section{ACKNOWLEDGMENTS}

The author would like to thank the Australian Research Council for their
financial support. He is indebted to his collaborators S.B.~Carr and 
B.F.~Gibson for many discussion and their helpful suggestions, and the 
Institute for Nuclear Theory at the University of Washington for 
creating the environment for the initial conversation between 
Carl Dover, Ben Gibson and the author that led to much of the work reported here.

\end{document}